\documentclass{emulateapj}

\slugcomment{To appear in ApJ Letters}

\shorttitle{Falling extrasolar giant planets}

\shortauthors{B. Levrard et al.}

\begin{document}

\title{Falling transiting extrasolar giant planets}

\author{B. Levrard\altaffilmark{1}, C. Winisdoerffer and G. Chabrier}
\affil{Ecole normale sup\'erieure de Lyon, Centre de Recherche Astrophysique de Lyon,
\\ Universit\'e de Lyon, 46 all\'ee d'Italie, F-69364 Lyon Cedex 07, France}


\altaffiltext{1}{also at IMCCE-CNRS UMR 8028, 77 Avenue Denfert-Rochereau, 75014, Paris. blevrard@ens-lyon.fr}

\begin{abstract}
We revisit the tidal stability of extrasolar systems harboring a transiting planet and demonstrate that, independently of any tidal model, none but one (HAT-P-2b) of these planets  has a tidal equilibrium state, which implies ultimately a collision of these objects with their host star. Consequently, conventional circularization and synchronization timescales cannot be defined because the corresponding states do not represent the endpoint of the tidal evolution. Using numerical simulations of the coupled tidal equations for the spin and orbital parameters of each transiting planetary system, we confirm these predictions and show that the orbital eccentricity and the stellar obliquity do not follow the usually assumed exponential relaxation but instead decrease significantly, reaching eventually a zero value, only during the final runaway merging of the planet with the star. The only characteristic evolution timescale of {\it all} rotational and orbital parameters is the lifetime of the system, which crucially depends on the magnitude of tidal dissipation within the star. These results imply that the nearly circular orbits of transiting planets and the alignment between the stellar spin axis and the planetary orbit are unlikely to be due to tidal dissipation. Other dissipative mechanisms, for instance interactions with the protoplanetary disk, must be invoked to explain these properties.
\end{abstract}


\section{Introduction}
Jupiter-like extra-solar planets have been detected transiting their parent
star at an unexpectedly small distance of less than 0.1 AU \citep[e.g.][]{Pon08}. Most of these
systems have nearly circular orbits \citep[e.g.][]{Pon08} and first measurements
of the sky-projected angle $\lambda$ between the stellar rotation axis and the planetary
orbital axis through the Rossiter-MacLaughlin effect
(for HD 209458, HD 149026, HD 189733, TrES-1, XO-1, HD 17156) indicate a nearly perfect
spin-orbit alignment \citep{Win05,Win07,Gau07,Nar07,Loe08,Coc08}.
These observational properties are commonly interpreted as an outcome of tidal dissipation between
the host star and the planet and these same effects are also believed to lead to synchronization of
the planetary and stellar rotation with the orbital motion. As a consequence, corresponding timescales
associated to these processes are usually evaluated by assuming an exponential relaxation towards equilibrium
parameters as obtained from any evolution perturbation calculation near an equilibrium state \citep[e.g.][]{Hut81}.
This leads to timescale estimates of spin-orbit alignment, synchronization and circularization which differ by
several orders of magnitude, ranging typically from $\sim 10^5$ yrs to a Hubble
time \citep[e.g.][]{Ras96b,Sas03,Dob04,Ogi08,Maz08}.  All these conclusions,
however, {\it implicitly assume the existence of such tidal equilibrium states.}

It has already been suggested that short-period planets could be unstable to tidal
 dissipation but these calculations were based on the assumption of the existence
of (unstable) tidal equilibrium states \citep{Ras96b,Dob04}, leading to an erroneous
application of the tidal stability criterion derived by \citet{Hut80}.
More recently, numerical simulations of the orbits of some transiting planets from the OGLE survey indicated a possible collapse with the host star but
the effect of tides raised by the star within the planet was ignored \citep{Pat04,Car07}. \citet{Jac08} noticed the importance of considering both tides raised by the star and the planet as well as the non-linear coupled evolution of the eccentricity and the orbital distance, but the global stability of the system and the additional coupling with the rotational evolution were not investigated.

In this Letter, we reconsider the stability of transiting extra-solar planets to tidal dissipation through theoretical and numerical
considerations and show that none but one of the transiting planets has a tidal equilibrium state. We investigate the consequences for the tidal evolution timescales of orbital (semi-major axis, eccentricity) and rotational (stellar obliquity, stellar and planetary rotational velocities) parameters, taking both tides raised by the planet and the star into account.

\section{Tidal stability of transiting planets}
A binary star-planet system that conserves the total
angular momentum $L_{\mathrm{tot}}$ but dissipates its energy
is known to dynamically evolve towards only two possible solutions \citep{Cou73,Hut80}.
On one hand, if $L_{\mathrm{tot}}<L_{\mathrm{c}}$, where $L_{\mathrm{c}}$ is the critical
angular momentum defined by
\begin{equation}
L_{\mathrm{c}}=4 \left[\frac{G^2}{27} \frac{M^{3}_{\star}
M_p^{3}}{M_{\star}+M_p}(C_p+C_{\star}) \right]^{1/4},
\end{equation}
where $M_p, C_p$ and $ M_{\star}, C_{\star}$ denote the masses and polar moments of inertia
of the planet and the star, respectively, and $G$ is the gravitational constant,
no equilibrium state exists and the system ultimately merges, {\it independently of
any tidal model}.
On the other hand, if $L_{\mathrm{tot}}>L_{\mathrm{c}}$, two equilibrium states exist that are characterized
 by the coincidence between equatorial and orbital planes, circularity of the orbit and
synchronization between rotational and orbital periods when no further dissipation occurs
\citep{Hut80}.
The furthest equilibrium orbital distance $a_1$ is stable while the closest $a_2$ is
unstable, with $a_{1} > a_{\mathrm{c}} > a_2$,
where $a_\mathrm{c}$ is the marginal equilibrium orbital distance for
$L_{\mathrm{tot}} = L_{\mathrm{c}}$, with
$a_\mathrm{c} \simeq \sqrt{3\,C_{\star}/M_p} \sim 0.064\,(M_{\mathrm{Jup}}/M_p)^{1/2}$ AU
for a planet orbiting a Sun-like star.
Therefore, exploring whether a system is stable or not, and deriving characteristic timescales, first requires to find out whether the condition $L_{\mathrm{tot}}<L_{\mathrm{c}}$ is satisfied or not.
Neglecting the spin of the planet, the total angular momentum of the system is given by the sum of the orbital angular momentum and the spin of the star: \begin{equation}
L_{\mathrm{tot}}=C_{\star} \omega_{\star}+ \frac{M_p\, M_{\star}}{\sqrt{M_p+M_{\star}}} \sqrt{G\,a\,(1-e^2)},
\end{equation}
assuming that the stellar obliquity is zero to maximize $L_{\mathrm{tot}}$,
where $e$ is the eccentricity, $\omega_{\star}$ is the rotational velocity of the star
and $C_{\star}=k \,M_{\star}\, R_{\star}^2$, where $k$ is set to the typical value 0.06
for centrally condensed stars with dominantly radiative interiors, characteristic
of Sun-like stars \citep{Cla89}. In order to investigate the fate of observed transiting planetary systems,
 we have computed the ratio $L_{\mathrm{tot}}/L_{\mathrm{c}}$ for each of them.  All the quantities relevant to our study
and their uncertainties have been collected from up-to-date published estimates (Table 1). The mean value of
$L_{\mathrm{tot}}/L_{\mathrm{c}}$ and its standard deviation have been estimated using a Monte-Carlo procedure
 by sampling the values of $M_{\star}$, $R_{\star}$, $M_p$, $a$ and $\omega_{\star}$, considered as normally
distributed around their most probable value. When the lower and upper error bars
for a variable are not the same,
the maximum of the two values has been considered as the standard deviation
of the distribution. When only an upper limit of the rotational velocity is
available, as for very slowly rotating stars, we considered a flat distribution
between 0 and the given upper limit.
Finally, we calculated the ratio $L_{\mathrm{tot}}/L_{\mathrm{c}}$
from a random set of the five variables and reproduced
this procedure $10^{6}$ times (Fig. 1).

For all but one case, the mean ratio $L_{\mathrm{tot}}/L_{\mathrm{c}}$ is  smaller than 1, indicating that none but one (HAT-P-2b) of the observed
transiting planets lies on a tidal equilibrium state and that they will ultimately all fall onto their parent star.
Hence, the exponential damping timescale originally put
forth by \citet{Ras96b} (as $a/\dot{a}$ or $e/\dot{e}$) is inappropriate because orbital circularization,
synchronization of the spins and spin-orbit alignment do not correspond to the endpoint of tidal evolution.
In the following, we define by timescale, the 
characteristic time required to observe a significant 
evolution of orbital or rotational parameters.

\begin{figure}[!t]
\epsscale{1.2}
\plotone{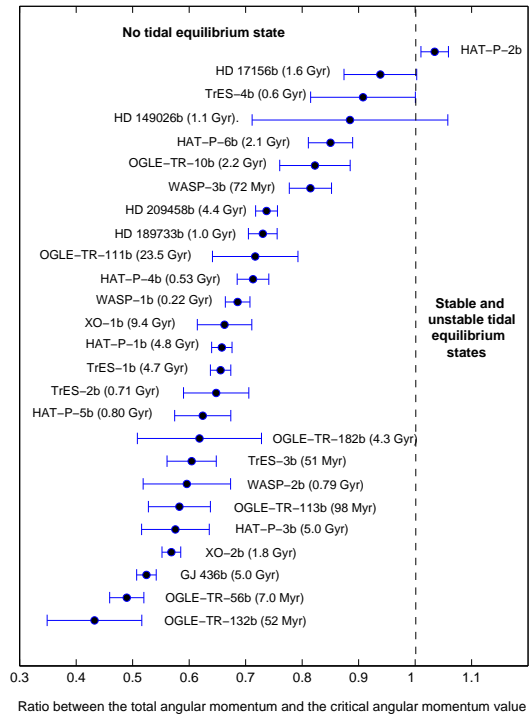}
\caption{List of transiting planets classified by increasing values of the ratio between the total angular momentum of the system
$L_{\mathrm{tot}}$ 
and the critical angular momentum $L_c$ (see text). The error bar indicates the standard deviation of the ratio distribution around its mean value. The time remaining for each planet to collide with its host star as obtained from our numerical computations with $Q'_{\star}=Q'_p=10^6$ (See Section 3) is indicated between brackets. Note that this time is strongly model-dependent. The HAT-P-2 system is tidally-unstable and very close to the marginally stable regime $a_1\simeq a_c$. Because its present orbital distance ($\sim$ 0.068 AU) is larger than the stable equilibrium value ($a_1 \sim$ 0.048 AU), the system will reach this state asymptotically.}
\end{figure}

\begin{figure*}[!t]
\epsscale{1.15}
\plottwo{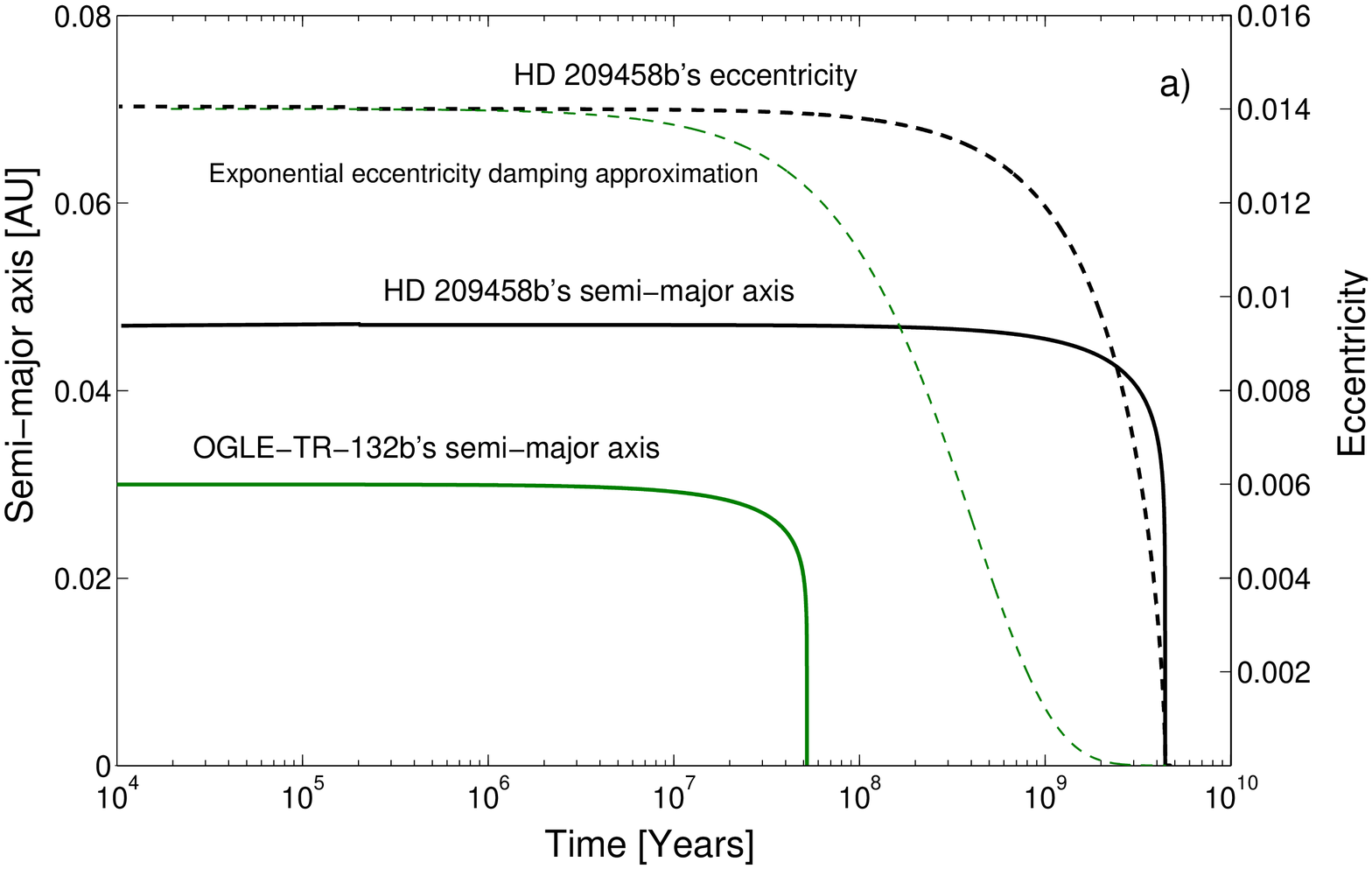}{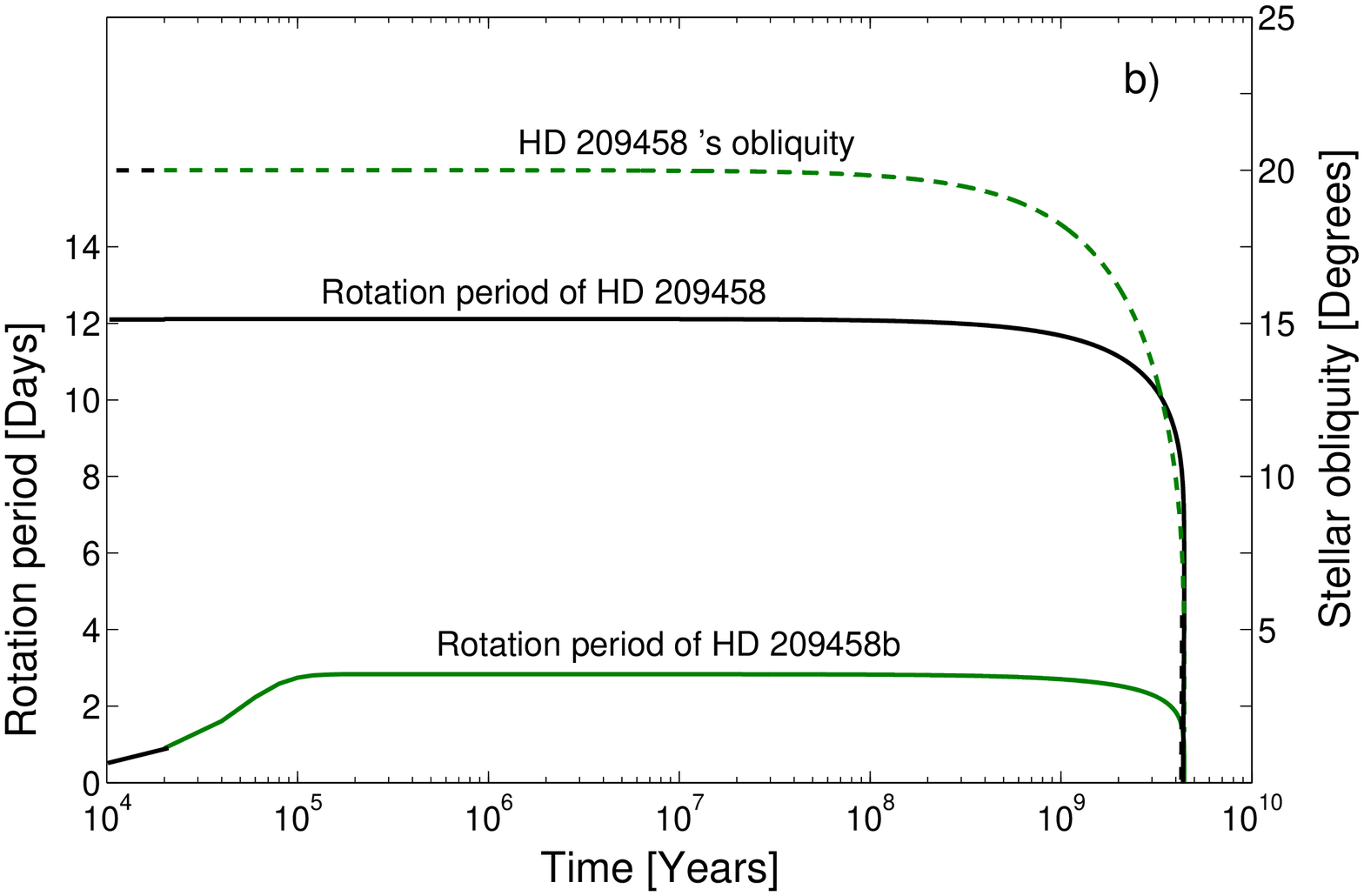}
\caption{Long-term integrations of orbital and rotational parameters of some transiting planets
and of their host star. Tidal quality parameters $Q'_{\star}$ and $Q'_p$ are initially set to $10^6$ but
vary with the semi-major axis as $Q_{\star}=Q_p=1/(n \Delta t)$  where $\Delta t$ is a constant tidal time lag. a) Semimajor axis of OGLE-TR-132b, HD 209458b  (solid lines) and eccentricity of HD 209458b (dashed line) as a function of time. The thin dashed line
is the usual exponential damping approximation of $e$ with a $4.1 \times 10^8$ yr relaxation time assuming that only tides raised on the planet are considered and stable tidal equilibrum states exist ; b) Rotational periods of HD 209458, HD 209458b (solid lines) and HD 209458's obliquity (dashed line) as a function of time. The initial rotational period of HD 209458b and HD 209458 stellar obliquity are arbitrarily set to half a day and $20^{\circ}$ to evaluate the respective timescales of their tidal evolution.}
\end{figure*}

\section{Timescales for the evolution of orbital and rotational parameters}
In order to determine the characteristic timescales of tidally-unstable extrasolar systems with no asymptotic equilibrium state, we have conducted long-term simulations of the fully coupled tidal equations for the orbital and spin parameters, taking tides raised both by the star and the planet into account. We follow the traditional ``viscous'' approach of the
equilibrium tide theory \citep{Dar08} which assumes a constant time lag for any frequency component between the tidally deformed surface of the gaseous envelope and the tidal perturbation, although alternative models are possible given our very limited
understanding of tidal processes in gaseous bodies \citep[e.g.][]{Ogi08}.
The rate of change in the semi-major axis is, at second order in eccentricity
\citep{Ner97}
\begin{eqnarray}
\frac{da}{dt}&=&\frac{6\,M_p\,R^5_{\star}}{Q'_{\star}\,M_{\star}\,a^4}\left[\left(1+\frac{27}{2}e^2\right)\,\cos \varepsilon
\,\omega_{\star}-(1+23\,e^2)n\right] \nonumber\\
&&\nonumber\\
&&+\frac{6\,M_{\star}\,R^5_p}{Q'_p\,M_p\,a^4}\left[\left(1+\frac{27}{2}e^2\right)\,
\omega_p-(1+23\,e^2)n\right],
\label{evol_a}
\end{eqnarray}
where $\omega_p$ is the planetary rotation velocity, $\varepsilon$ is the stellar obliquity and $Q'_{\star}$ (resp. $Q'_p$) the ratio between the present annual stellar (resp. planetary) tidal quality factor $Q_{\star}$ (resp. $Q_p$) and the tidal Love number of degree 2 $k_{2,\star}$ (resp. $k_{2,p}$). The first term in Equation (\ref{evol_a}) reflects the effects of tides within the star while the
effects of the tide raised by the star within the planet are reflected in the second term. We assume that the planetary obliquity is zero. Similar equations for the evolution of $e$, $\varepsilon$, $\omega_p$ and $\omega_{\star}$ can be found
in \citet{Hut81}.  Since the orbits of transiting planets are nearly edge-on, the angle $\lambda$ is related to the stellar obliquity $\varepsilon$ by $\cos \varepsilon \simeq \cos \lambda \sin i_{\star}$, where $i_\star$ is the inclination of the stellar rotation
axis relative to the sky plane \citep{Win05}. In order to evaluate the typical timescale for the tidal evolution of the projected spin-orbit separation, the initial stellar obliquity was then modified and set to some arbitrary non-zero values. We numerically integrated the complete set of tidal equations forward in time
using {\it present} values as initial conditions for each transiting planetary
system and setting the same initial value $10^6$ for $Q'_{\star}$ and $Q'_p$.

The typical evolution of the orbital parameters is shown in Figure 2a for an eccentric
 planet (HD 209458b) and for the ``most unstable''  transiting planet OGLE-TR-132b which
 lies on a circular orbit. As expected, the evolution of the orbits completely departs
from an exponential relaxation, characteristic of the existence of stable or unstable
tidal equilibrium states. Both  semi-major axis evolve slowly until the planets abruptly collide with their parent star (within $\lesssim$ 10 Myr and $\lesssim$ 0.5 Gyr, respectively), reaching a zero value in $\sim$ 52 Myr and 4.3 Gyr, respectively. We verified that the same issue holds true for the other unstable transiting planetary systems and the time required for each planet to reach its host star is indicated in Figure 1 for an initial zero stellar obliquity. It typically ranges from 7 Myr (OGLE-TR-56) to 23.5 Gyr (OGLE-TR-111).  Similarly (see Fig. 2a), HD 209458b's eccentricity remains also nearly constant until an ultimate runaway decrease occurs associated to the orbital shrinkage.
For comparison, we evaluate for this planet the timescales derived from the usual exponential tidal
damping solution, based on the erroneous assumption that transiting planets are close
to their tidal equilibrium state : when only tides raised by the star (resp. planet)
on the planet (resp. star) are considered, $\tau_{e,p}=(2/21)\times(Q'_p/n)\,(a/R_p)^5\,(M_p/M_{\star})
\simeq 0.41$ Gyr 
(resp. $\tau_{e,\star} = (2/21)\times (Q'_{\star}/n)\,(a/R_{\star})^5\,(M_{\star}/M_p) \simeq 15$ Gyr), 
more than one order of magnitude smaller (resp. three times larger) than 
the afore correctly calculated remaining time before the eccentricity reaches zero.


The evolution of the rotational parameters is illustrated for the HD 209458 system in Figure 2b.
Despite the global tidal instability, the planetary rotation velocity is first pseudo-synchronized with the orbital motion $n$ such that $\omega_p \sim (1+6\,e^2)\,n$ over the same timescale $\tau_{ps} \sim (C_p \,n\,Q'_p\,a^6)/(3\,G\,M^2_{\star}\,R^5_p)$, as
the one estimated when tidal equilibrium states exist \citep{Hut81}. It is close to
$\sim 10^5$~yr for typical HD 209458b's parameters, much smaller than the orbital decay
 timescale. This stems from the fact that
the angular momentum associated with the planet's rotation is much smaller
 than the orbital angular momentum.
However, Figure 2b clearly shows that it only corresponds to a {\it temporary} state
because the ultimate orbital collapse causes the orbital mean motion to diverge and
then the rotational period of the planet to spin up dramatically.
The evolution of HD 209458's rotational period can be easily understood in the framework
of classical tidal theory
\citep[e.g.][]{Hut81, Pat04, Car07}: because HD 209458 rotates more slowly than its
 orbital period, the tidal torque raised by the planet onto the star yields the star
to spin up with time until a runaway spin-up occurs, due to the conservation of
momentum. Hence, HD 209458 is expected to never reach synchronization, contrary
to current assumptions.
The stellar obliquity follows the same trends and tidal dissipation acts
to coplanarize the orbit and the stellar equator {\it only} during the
 final runaway phase of the orbital decay. Figure 2 shows that, except for
 the existence of a rapid pseudo-synchronization timescale for the planetary
rotation, the timescale for the tidal evolution of {\it all} the orbital and
rotational parameters is the time required for the semi-major axis
to shrink to zero, {\it i.e.} the remaining lifetime of the system,
 that has to be evaluated.

There is no simple analytical solution for eqn.(\ref{evol_a}) but some reasonable simplifications are possible. Indeed, once the planetary spin has rapidly reached its pseudo-equilibrium state within $\sim 10^{5}-10^{6}$ yrs, any further exchange of angular momentum between the planet's rotation and its orbit only occurs through radial tides due to a non-zero eccentricity.
Assuming that the planet's orbital period is short compared with the star's rotation period, which is true for all unstable transiting planets, eqn.(\ref{evol_a}) shows that the rate of orbital decay driven by planetary tides becomes smaller than the contribution of tides raised by the planet on the star if the relation:
\begin{equation}
e <   \frac{M_{p}}{M_{\star}} \sqrt{\frac{2\,Q'_{p}}{7\,Q'_{\star}}}  \left(\frac{R_{\star}}{R_{p}} \right)^{5/2}
\label{e_max}
\end{equation}
is verified.  This condition is fulfilled for all unstable transiting planetary systems (considering $Q'_{p} \simeq Q'_{\star}$)
\footnote{except for GJ 436 for which $\sqrt{2/7}(M_{p}/M_{\star})\times(R_{\star}/R_{p})^{5/2} \sim 0.08 \lesssim e \sim 0.14$.}.
In this context,  eqn.(\ref{evol_a}) can be integrated, giving the remaining time for the planet to
reach its host star from its initial orbital distance $a$ (at small eccentricity):
\begin{equation}
\tau_a \simeq \frac{1}{48}\frac{Q'_{\star}}{n}\,\left(\frac{a}{R_{\star}}\right)^5\,\left(\frac{M_{\star}}{M_p}\right) \,,
\label{timing_a}
\end{equation}
which only depends on tidal dissipation within the star, not within the planet \footnote{
Using a tidal model for which the quality factor is independent of the
tidal frequency yields a similar value (the factor 1/48 must be replaced by 2/39)
\citep{Pat04}
}. 
We stress, however, that such timescales are very uncertain, because of the
large uncertainties in the estimate of the $Q'_{\star}$ value (typically
$10^5 < Q'_{\star} < 10^{10}$) \citep[e.g.][]{Pat04}.
Other tidal models involving dissipation by turbulent viscosity in the
convective zone provide estimates of the remaining lifetime several orders
of magnitude larger \citep{Ras96b,Sas03,Pat04}.
We found the values obtained with eqn.(\ref{timing_a}) to be in good agreement with those determined from numerical computations for each transiting planet, except for HD 17156 and GJ 436 which have significant
eccentricities, leading to a net decrease of the system lifetime due to enhanced tidal interactions.
A non-zero stellar obliquity also causes the lifetime of the system to be slightly reduced.

\section{Discussion and conclusions}

Based on rigorous arguments and calculations, we have demonstrated that none but one of the discovered transiting extra-solar planet has a tidal equilibrium state, implying a collapse with their host star. The lack of tidal equilibrium states implies that all the orbital and rotational parameters evolve over the {\it same}, only relevant timescale which corresponds to the lifetime of the system. We stress that these results are independent of the tidal model and can probably be extended to most other close-in giant planets detected by radial velocities.

Nevertheless, the exact evaluation of such a timescale strongly depends on the nature of the tidal processes
which remain poorly constrained \citep[see e.g.]
[]{Zah77, Ras96b, Sas03, Pat04}. The very existence of the currently
 observed transiting planets suggests that the lifetime of these systems is at least equal to
the age of the systems. An alternative possibility is that one or several undetected planetary companions maintain the stability of the present orbit if the planetary orbits are in resonance.
Finally, because our results show that both the orbital eccentricity and the stellar obliquity undergo a substantial decay only during the ultimate runaway orbital collapse, this implies that tidal dissipation {\it in the star and in the planet} has probably not played a dominant role in the current observations of nearly-circular orbits and spin-orbit alignments of planetary transiting systems, unless each of them is now precisely in this final rapid merging state, a rather unlikely possibility.
These results bear important consequences for our understanding of planet formation, migration and planet-disk interaction.
They suggest that either the systems were formed with a nearly circular orbit and a stellar spin nearly aligned with the orbital angular momentum, or other processes such as gravitational interactions with the protoplanetary disk have dissipated the initial eccentricity \citep{Moo08}. In contrast, planet-planet scattering during the early stages of planet formation \citep[e.g.][]{Cha08} or gravitational perturbations by a companion star could have randomized spin-orbit alignment and/or produced large eccentricities \citep{Ras96a}.
For sake of simplicity, we have considered that the total angular momentum of the system is constant over time. In reality, angular momentum of the star is continually extracted by a magnetized stellar wind, so that the ratio $L_{\mathrm{tot}}/L_{\mathrm{c}}$ should decrease with time. This
indicates that binary systems close to the marginally stable regime like HAT-P-2b (for which $a_1\simeq a_c$) may eventually become unstable to tidal dissipation. One also may argue that tidal dissipation is mostly effective in stellar convective layers \citep{Zah77} so that only the moment of inertia of the outer convective envelope, about $\sim 5\%$ of the total moment of inertia $C_{\star}$ for a Sun-like star,
should be considered both in eqns (1) and (2). In that case, the system « convective layers + planet » is not isolated
because the differential rotation between the convective and radiative zones generates an extra dissipative torque until radiative and convective zones are synchronized. Although this may affect the evolution timescale of the system, the final conclusion should remain the same.
We postulate that satellite missions like CoRot or Kepler should detect transiting planets on stable equilibrium orbits, with $L_{tot}>L_c$ and $a>a_2$, further away from the tidal instability limit.


\acknowledgments
The authors are endebted to P. Robutel for useful discussions. This research is supported by the CNRS ``Programme National de Plan\'etologie".


\newpage

\begin{figure}[!t]
\epsscale{1.2}
\plotone{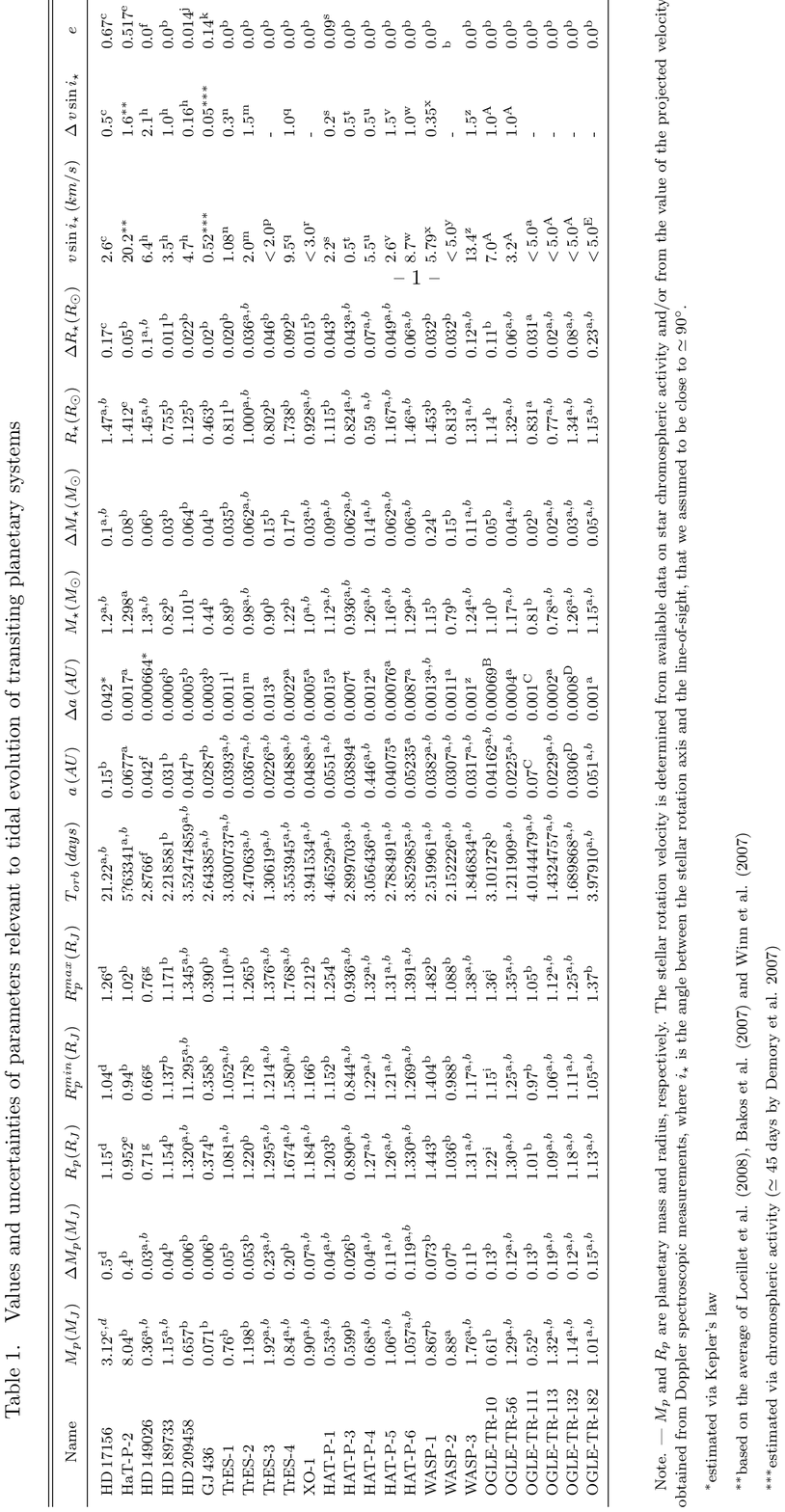}
\end{figure}

\end{document}